\documentclass[aps,prd,twocolumn,citeautoscript,superscriptaddress,nofootinbib]{revtex4-1}  
\usepackage{amsmath,amssymb,bm,bbm} 
\usepackage{graphicx}  
\usepackage{color} 
\usepackage[colorlinks=true]{hyperref}  
\hypersetup{
    bookmarks=true,         
    unicode=false,          
    pdftoolbar=true,        
    pdfmenubar=true,        
    pdffitwindow=false,     
    pdfstartview={FitH},    
    pdftitle={Z2 topological order near the Neel state of the square lattice antiferromagnet},    
    pdfauthor={Shubhayu Chatterjee, Subir Sachdev and Mathias Scheurer},     
    pdfsubject={},   
    pdfcreator={},   
    pdfproducer={}, 
    pdfkeywords={} {} {}, 
    pdfnewwindow=true,      
    colorlinks=true,       
    linkcolor=magenta, 
    citecolor=blue,        
    filecolor=magenta,      
    urlcolor=blue           
} 

\usepackage{amsfonts}
\usepackage{upgreek}
\usepackage{slashed}
\usepackage{latexsym}
\usepackage{dsfont}
\usepackage{todonotes}

\newcommand{\beq}{\begin{equation}}
\newcommand{\eeq}{\end{equation}}
\def\bea{\begin{eqnarray}}
\def\eea{\end{eqnarray}}
\def \k{{\bm k}}
\def \Q{{\bm K}}

\newcommand{\nn}{\nonumber \\}
\renewcommand{\vec}[1]{\boldsymbol{#1}}

\newcommand{\diff}{d}
\newcommand{\pdagger}{{\phantom{\dagger}}}
\newcommand{\equref}[1]{Eq.~(\ref{#1})}
\usepackage{braket}

\begin{document}

\title{Intertwining topological order and broken symmetry\\ in a theory of fluctuating spin density waves}

\author{Shubhayu Chatterjee}
\affiliation{Department of Physics, Harvard University, Cambridge MA 02138, USA}

\author{Subir Sachdev}
\affiliation{Department of Physics, Harvard University, Cambridge MA 02138, USA}
\affiliation{Perimeter Institute for Theoretical Physics, Waterloo, Ontario, Canada N2L 2Y5}

\author{Mathias S.~Scheurer}
\affiliation{Department of Physics, Harvard University, Cambridge MA 02138, USA}

\date{\today}

\begin{abstract}
The pseudogap metal phase of the hole-doped cuprate superconductors has two seemingly unrelated characteristics: a gap in the electronic spectrum
in the `anti-nodal' region of the square lattice Brillouin zone, and discrete broken symmetries. 
We present a SU(2) gauge theory of quantum fluctuations of magnetically ordered states which appear 
in a classical theory of square lattice antiferromagnets, in a spin density wave mean field theory of the square lattice Hubbard model, and in a $\mathbb{CP}^1$
theory of spinons.
This theory leads to metals with an antinodal gap, and topological order which intertwines with precisely the observed broken symmetries.
\end{abstract}
\maketitle

A remarkable property of the pseudogap metal of the hole-doped cuprates is that it 
does not exhibit a `large' Fermi surface of gapless electron-like  quasiparticles excitations, \emph{i.e.}~the size of the Fermi surface is smaller than expected from the classic Luttinger theorem of Fermi liquid theory \cite{Luttinger}. 
Instead it
has a gap in the fermionic spectrum near the `anti-nodal' points
($(\pi, 0)$ and $(0, \pi)$) of the square lattice Brillouin zone.
Gapless fermionic excitations appear to be present only along the diagonals of the Brillouin zone (the `nodal' region).
One way to obtain such a Fermi surface reconstruction is by a broken translational symmetry. However, there is no sign of broken translational symmetry over a wide intermediate temperature range \cite{Keimer15}, and also at low temperatures and intermediate doping \cite{LTCP15}, over which the pseudogap is present. With full translational symmetry, violations of the Luttinger theorem require the presence of topological order \cite{FFL,TSMVSS04,APAV04}.

A seemingly unrelated property of the pseudogap metal is that it exhibits discrete broken symmetries, which preserve translations, over roughly the same region of the phase diagram over which there
is an antinodal gap in the fermionic spectrum. 
The broken symmetries include lattice rotations, interpreted in terms of an Ising-nematic order \cite{Ando02,Hinkov597,2010Natur.463..519D,2010Natur.466..347L}, and one or both of inversion and time-reversal symmetry breaking \cite{Bourges06,2008arXiv0805.2959L,2008PhRvL.100l7002X,2010Natur.468..283L,2014PhRvL.112n7001L,2015NatCo...6E7705M,2016arXiv161108603Z}, usually interpreted in terms of Varma's current loop order \cite{SimonVarma}.
Luttinger's theorem implies that none of these broken 
symmetries can induce the needed fermionic gap by themselves. 

The co-existence of the antinodal
gap and the broken symmetries can be explained by intertwining them \cite{SSNR91,2013PhRvB..87n0402B,SSSC17}, {\it i.e.}~by exploiting flavors of topological order which are tied to specific broken symmetries. Here we show that the needed flavors
appear naturally in several models appropriate to the known cuprate electronic structure.

\begin{figure*}
\begin{center}
\includegraphics[width=6.5in]{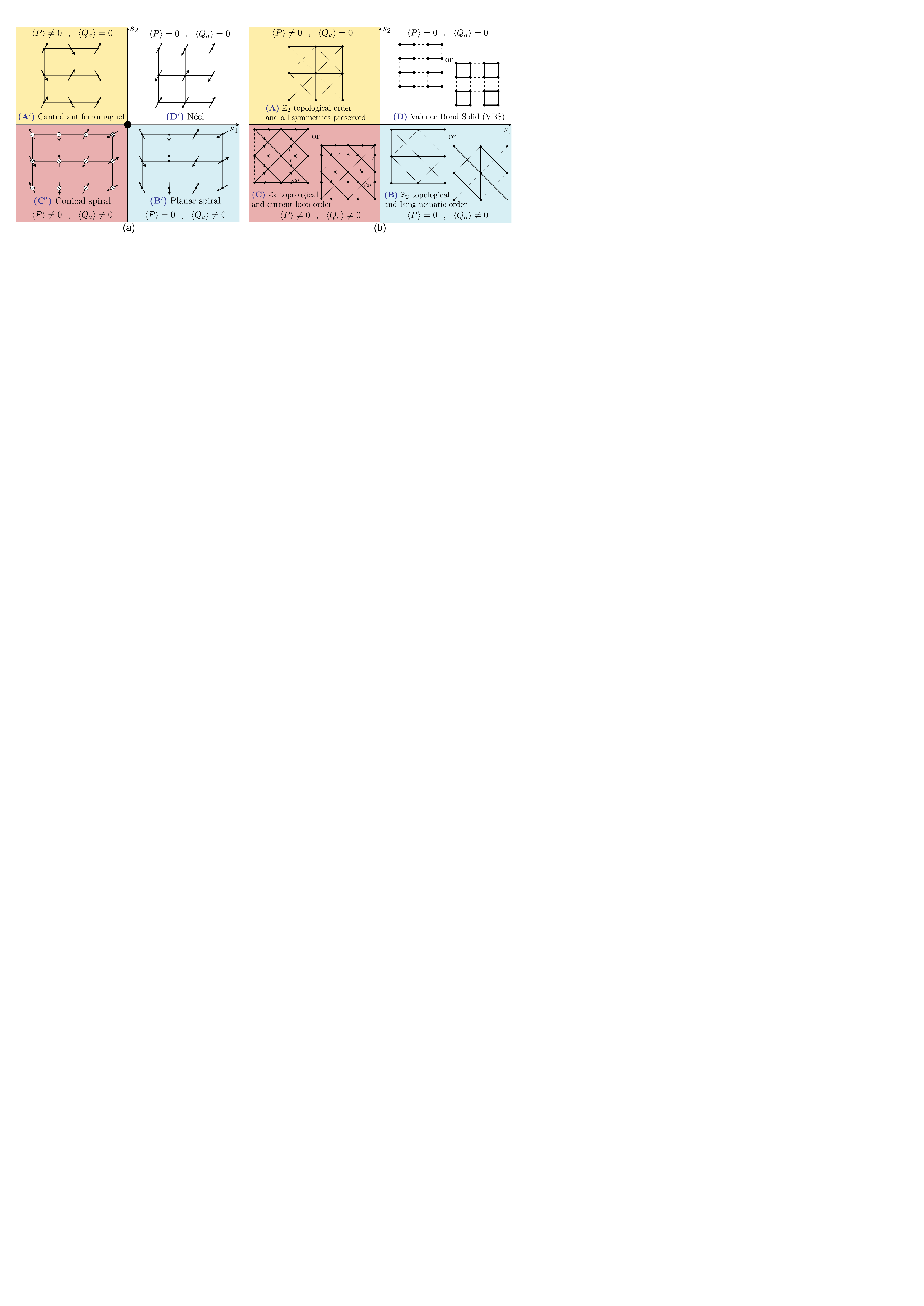}
\end{center}
\caption{(a) Schematics of the magnetically ordered states obtained in the classical antiferromagnet, and in the spin density wave theory of the Hubbard model. (b) Corresponding states obtained after quantum fluctuations restore spin rotation symmetry. Phase D has U(1) topological order in the metal, but is unstable to the appearance of VBS order in the insulator. The crossed circles in phase C$^\prime$ indicate a canting of the spins into the plane. The labels $s_1$, $s_2$, $P$, $Q_a$ refer to the $\mathbb{CP}^1$ theory: the phases in (a) are obtained for small $g$, and those in (b) for large $g$.}
\label{fig:pdiags}
\end{figure*}

We consider quantum fluctuations of magnetically ordered states found in two different computations: a classical theory of frustrated, insulating
antiferromagnets on the square lattice, and a spin density wave theory of metallic states of the square lattice Hubbard model. The types of magnetically ordered
states found are sketched in Fig.~\ref{fig:pdiags}a. The quantum fluctuations of these states are described by a SU(2) gauge theory, and this leads to the loss
of magnetic order, and the appearance of phases with topological order and an anti-nodal gap in the fermion spectrum. We find that the topological order
intertwines with precisely the observed broken discrete symmetries, as shown in Fig.~\ref{fig:pdiags}b.
We further show that the same phases are
also obtained naturally in a $\mathbb{CP}^1$ theory of bosonic spinons supplemented by Higgs fields conjugate to long-wavelength spinon pairs.

\noindent
{\bf Magnetic order:}
We examine states in which the electron spin $\hat{\bm S}_i$ on site $i$ of the square lattice, at position $\vec{r}_i$, has the expectation value
\bea
\left\langle \hat{\bm S}_i \right\rangle &=& N_0 \left[ \cos \left( \vec{K} \cdot \vec{r}_i \right) \cos(\theta) \, \hat{\bm e}_x + 
\sin \left( \vec{K} \cdot \vec{r}_i \right) \cos(\theta) \, \hat{\bm e}_y \right. \nn
&+& \left.  \sin (\theta)\, \hat{\bm e}_z\right].
\label{spinansatz}
\eea
The different states we find are (see Fig.~\ref{fig:pdiags}a) (D$^\prime$) a N\'eel state with collinear antiferromagnetism at wavevector $(\pi, \pi)$, with
$\vec{K}= (\pi, \pi)$, $\theta=0$; (A$^\prime$) a canted state, with $(\pi,\pi)$ N\'eel order co-existing with a ferromagnet moment perpendicular to the N\'eel order, with $\vec{K}= (\pi, \pi)$, $0<\theta<\pi/2$,
(B$^\prime$) a planar spiral state, in which the spins precess at an incommensurate wavevector $\vec{K}$ with $\theta=0$; (C$^\prime$) 
a conical spiral state, which is a planar spiral accompanied by a ferromagnetic moment perpendicular to the plane of the spiral \cite{Wiesendanger12} with
$\vec{K}$ incommensurate, $0<\theta<\pi/2$.

\begin{figure*}
\begin{center}
\includegraphics[width=6.5in]{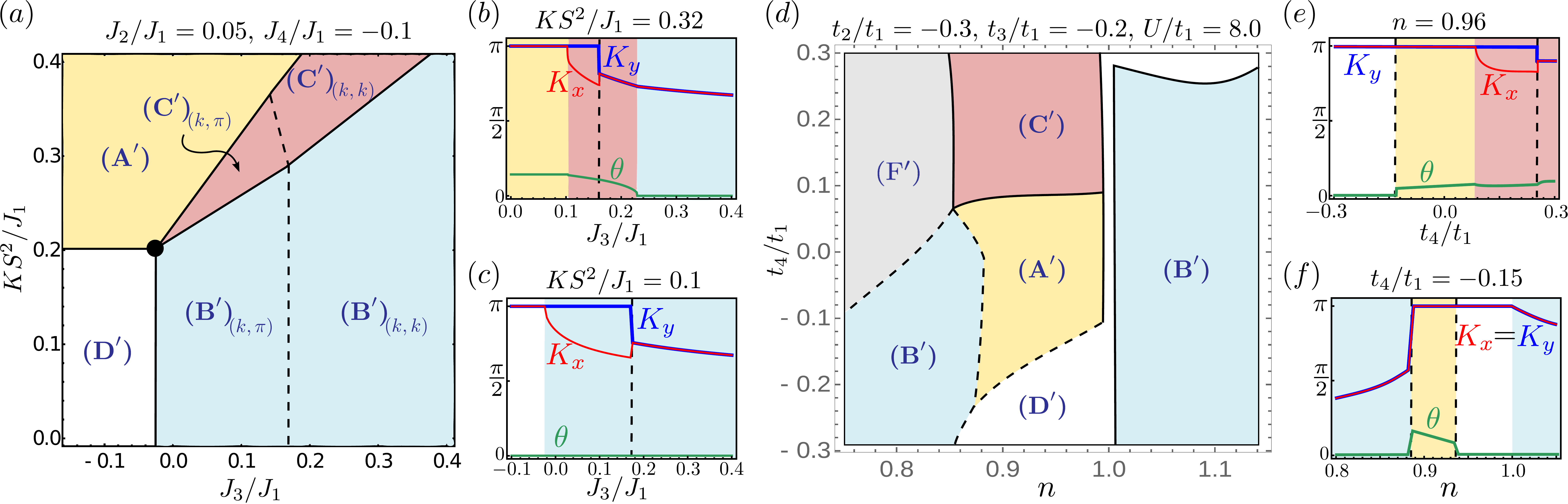}
\end{center}
\caption{(a) Phase diagram of $\mathcal{H}_J$, for a spin $S$ model in the classical limit $S \rightarrow \infty$, exhibiting all phases of Fig.~\ref{fig:pdiags}a. The subscript of the labels $(\text{B}')$ and $(\text{C}')$ indicates the wavevector $\vec{K} = (K_x,K_y)$ of the spiral. Note that the phases A$^\prime$, C$^\prime$, B$^\prime$, D$^\prime$ meet at a multicritical point, just as in Fig~\ref{fig:pdiags}a. (b) and (c) show $K_x$, $K_y$, and the canting angle $\theta$ along two different one-dimensional cuts of the phase diagram in (a). The phase diagram resulting from the spin-density wave analysis of the Hubbard model (\ref{HubbardModel}) can be found in (d).  Besides an additional ferromagnetic phase, denoted by $(\text{F}')$, we recover all the phases of the classical phase diagram in (a). Part (e) and (f) show one-dimensional cuts of the spin-density wave phase diagram. In all figures, solid (dashed) lines are used to represent second (first) order transitions.}
\label{fig:magorder}
\end{figure*}

First, we study the
square lattice spin Hamiltonian with near-neighbor antiferromagnetic exchange interactions $J_p>0$, and ring exchange $K$ \cite{MGY88,GSH88,Chubukov92,Lauchli05,Uhrig12}:
\bea
\mathcal{H}_J &=& \sum_{i<j}  J_{ij} \, \hat{\bm S}_i \cdot \hat{\bm S}_{j} + 2 K \sum_{\setlength\unitlength{0.3pt}
\begin{picture}(35,40)
\put(12,4){\line(1,0){20}}
\put(12,4){\line(0,1){20}}
\put(12,24){\line(1,0){20}}
\put(32,4){\line(0,1){20}}
\put(34,0){\tiny $i$}
\put(0,0){\tiny $j$}
\put(0,27){\tiny $k$}
\put(31,27){\tiny $\ell$}
\end{picture}} \Bigl[ (\hat{\bm S}_i \cdot \hat{\bm S}_{j})(\hat{\bm S}_k \cdot \hat{\bm S}_{\ell}) \nn
&+& (\hat{\bm S}_i \cdot \hat{\bm S}_{\ell})(\hat{\bm S}_k \cdot \hat{\bm S}_{j})
- (\hat{\bm S}_i \cdot \hat{\bm S}_{k})(\hat{\bm S}_j \cdot \hat{\bm S}_{\ell})\Bigr]\,. \label{spinH}
\eea
$J_{ij} = J_p$ when $i,j$ are $p$'th nearest neighbors,
and we only allow $J_p$ with $p=1,2,3,4$ non-zero. The classical ground states are obtained by minimizing $\mathcal{H}_J$
over the set of states in Eq.~(\ref{spinansatz}); results are shown
in Fig.~\ref{fig:magorder}a-c. We find the states A$^\prime$, B$^\prime$, C$^\prime$, D$^\prime$, all of which
meet at a multicritical point, just as in the schematic phase diagram in Fig.~\ref{fig:pdiags}a. A semiclassical theory of quantum fluctuations about these states, starting from the N\'eel state, appears in Appendix~A.

For metallic states with spin density wave order, we study the Hubbard model 
\beq
\mathcal{H}_U = - \sum_{i<j, \alpha} t_{ij} c_{i,\alpha}^\dagger c_{j,\alpha}^{\vphantom\dagger}
-\mu \sum_{i, \alpha} c_{i, \alpha}^\dagger c_{i,  \alpha}^{\vphantom\dagger}
+  U \sum_i \hat{n}_{i, \uparrow} \hat{n}_{i, \downarrow} \label{HubbardModel}
\eeq
of electrons $c_{i,\alpha}$, with $\alpha = \uparrow,\downarrow$ a spin index, $t_{ij} = t_p$ when $i,j$ are $p$'th nearest neighbors,
and we take $t_p$ with $p=1,2,3,4$ non-zero. $U$ is the on-site Coulomb repulsion, and $\mu$ is the chemical potential. The electron density, $\hat{n}_{i, \alpha} 
\equiv c_{i,\alpha}^\dagger c_{i,\alpha}^{\vphantom\dagger}$, while the electron spin $\hat{\bm S}_i \equiv (1/2) 
c_{i,\alpha}^\dagger {\bm \sigma}_{\alpha\beta} c_{i,\beta}^{\vphantom\dagger}$, with ${\bm \sigma}$ the Pauli matrices.
We minimized $\mathcal{H}_U$ over the set of free fermion Slater determinant states obeying Eq.~(\ref{spinansatz}), while
maintaining uniform charge and current densities; results are illustrated in Fig.~\ref{fig:magorder}d-f, and details appear in Appendix~B. Again, note the appearance of the magnetic orders A$^\prime$, B$^\prime$, C$^\prime$, D$^\prime$,
although now these co-exist with Fermi surfaces and metallic conduction.

\noindent
{\bf SU(2) gauge theory:}
We describe quantum fluctuations about states of $\mathcal{H}_U$ obeying Eq.~(\ref{spinansatz}) by transforming the electrons to a
rotating reference frame by a SU(2) matrix $R_i$ \cite{SS09}
\beq
\left( \begin{array}{c} c_{i, \uparrow} \\ c_{i, \downarrow} \end{array} \right) = R_i \left( \begin{array}{c} \psi_{i,+} \\ \psi_{i,-} \end{array} \right), \quad\quad R_i^\dagger R_i = R_i R_i^\dagger = \mathds{1}.
\label{R}
\eeq
The fermions in the rotating reference frame are spinless `chargons' $\psi_{s}$, with $s=\pm$, carrying the electromagnetic charge. In the same manner, the transformation of the electron spin operator $\hat{\bm S}_i$ to the rotating reference frame is proportional to the  `Higgs' field ${\bm H}_i$ \cite{SS09},
\begin{equation}
{\bm \sigma} \cdot {\bm H}_i \propto  R_i^\dagger \, {\bm \sigma} \cdot \hat{\bm S}_i \, R_i \label{H}.
\end{equation}
The new variables, $\psi$, $R$, and ${\bm H}$ provide a formally redundant description of the physics of $\mathcal{H}_U$
as all observables are invariant under a SU(2) gauge transformation $V_i$ under which
\beq
\begin{array}{c}
R_i \rightarrow R_i \, V_i^\dagger \\
 {\bm \sigma} \cdot {\bm H}_i \rightarrow V_i \, {\bm \sigma} \cdot {\bm H}_i\, V_i^\dagger ~
 \end{array}
 \left( \begin{array}{c} \psi_{i,+} \\ \psi_{i,-} \end{array} \right) \rightarrow V_i \left( \begin{array}{c} \psi_{i,+} \\ \psi_{i,-} \end{array} \right) 
 , 
 \label{GaugeTrafo}
\eeq
while $c_i$ and $\hat{\bm S}_i$ are gauge invariant. The action of the SU(2) gauge transformation $V_i$, should
be distinguished from the action of global SU(2) spin rotations $\Omega$ under which
\beq
\begin{array}{c}
R_i \rightarrow \Omega \, R_i \\ 
{\bm \sigma} \cdot \hat{\bm S}_i \rightarrow \Omega \, {\bm \sigma} \cdot \hat{\bm S}_i\, \Omega^\dagger
\end{array}
~
\left( \begin{array}{c} c_{i\uparrow} \\ c_{i\downarrow} \end{array} \right) \rightarrow
\Omega \left( \begin{array}{c} c_{i\uparrow} \\ c_{i\downarrow} \end{array} \right),
\eeq
while $\psi$ and ${\bm H}$ are invariant.

In the language of this SU(2) gauge theory \cite{SS09,DCSS15b}, the phases with magnetic order obtained above appear 
when both $R$ and ${\bm H}$
are condensed. We may choose a gauge in which $\langle R \rangle \propto \mathds{1}$, and so the orientation
of the ${\bm H}$ condensate is the same as that in Eq.~(\ref{spinansatz}),
\bea
\left\langle {\bm H}_i \right\rangle &=& H_0 \Bigl[ \cos \left( \vec{K} \cdot \vec{r}_i \right) \cos(\theta) \, \hat{\bm e}_x + 
\sin \left( \vec{K} \cdot \vec{r}_i \right) \cos(\theta) \, \hat{\bm e}_y \nn
&+&  \sin (\theta)\, \hat{\bm e}_z\Bigr].
\label{higgsansatz}
\eea

We can now obtain the phases of $\mathcal{H}_U$ with quantum fluctuating spin density wave order, (A,B,C,D) shown in Fig.~\ref{fig:pdiags}b, in a simple step: the quantum fluctuations lead to fluctuations in the orientation of the local magnetic order, and so remove the $R$ condensate leading to $\langle R \rangle = 0$. The
Higgs field ${\bm H}_i$ retains the condensate in Eq.~(\ref{higgsansatz}) indicating that the magnitude of the local order is non-zero. In such a phase, spin rotation invariance is maintained with $\langle \hat{\bm S} \rangle = 0$, but the SU(2) gauge group has been `Higgsed'
down to a smaller gauge group which describes the topological order \cite{NRSS91,SSNR91,Wen91,Bais92,MMS01,Hansson04}. The values of $\theta$ and $\vec{K}$ in phases (A,B,C,D) obey the same constraints as the corresponding magnetically ordered phases (A$^\prime$, B$^\prime$,
C$^\prime$, D$^\prime$). In phase D, the gauge group is broken down to U(1), and there is a potentially gapless emergent `photon'; in an insulator, monopole condensation drives confinement and the appearance of VBS order,
but the photon survives in a metallic, U(1) `algebraic charge liquid' (ACL) state \cite{KKSS08} (which is eventually unstable to fermion pairing and superconductivity \cite{Mross15}). The remaining phases A,B,C
have a non-collinear configuration of $\langle {\bm H}_i \rangle$ and then only $\mathbb{Z}_2$ topological order survives \cite{SSNR91}: such states are ACLs with stable, gapped, `vison' excitations carrying $\mathbb{Z}_2$ gauge flux which cannot be created singly by any local operator. 
Phase A breaks no symmetries, phase B breaks lattice rotation symmetry leading to Ising-nematic order \cite{NRSS91,SSNR91}, and
phase C has broken time-reversal and mirror symmetries (but not their product), leading to current loop order.
All the 4 ACL phases (A,B,C,D) may also become
`fractionalized Fermi liquids' (FL*) \cite{FFL,TSMVSS04} by formation of bound states between the chargons and $R$; the FL* states
have a Pauli contribution to the spin susceptibility from the reconstructed Fermi surfaces.

The structure of the fermionic excitations in the phases of Fig.~\ref{fig:pdiags}b, and the possible broken
symmetries in the $\mathbb{Z}_2$ phases, can be understood from an effective Hamiltonian for the chargons. As described in Appendix~C, a Hubbard-Stratonovich transformation on $\mathcal{H}_U$, followed 
by the change of variables in Eqs.~(\ref{R}) and (\ref{H}), and a mean field decoupling leads to 
\bea
\mathcal{H}_\psi &=& - \sum_{i<j, s} t_{ij} Z_{ij} \psi_{i,s}^\dagger \psi_{j,s}^{\vphantom\dagger} 
-\mu \sum_{i, s} \psi_{i,s}^\dagger \psi_{i, s}^{\vphantom\dagger} \nn
&-& \sum_{i,s,s'} {\bm H}_i \cdot \psi_{i,s}^\dagger {\bm \sigma}_{ss'} \psi_{i,s'}^{\vphantom\dagger} \,. \label{ChargonHam}
\eea
The chargons inherit their hopping from the electrons, apart from a renormalization factor $Z_{ij}$, and experience
a Zeeman-like coupling to a local field given by the condensate of ${\bm H}$: so the Fermi surface of $\psi$
reconstructs in the same manner as the Fermi surface of $c$ in the phases with conventional spin density wave order. 
Note that this happens here even though translational symmetry is fully preserved in all gauge-invariant observables;
the apparent breaking of translational symmetry in the Higgs condensate in Eq.~(\ref{higgsansatz}) does not transfer to any gauge invariant observables, showing how the Luttinger theorem can be violated by the topological order \cite{FFL,TSMVSS04,APAV04} in Higgs phases. However, other symmetries are broken in gauge-invariant observables: Appendix~C examines
bond and current variables, which are bilinears in $\psi$, and finds that they break symmetries in the phases
B and C noted above.

\noindent
{\bf $\mathbb{CP}^1$ theory:}
We now present an alternative description of all 8 phases in Fig.~\ref{fig:pdiags} starting from the popular $\mathbb{CP}^1$ theory of quantum antiferromagnets. In principle (as we note below, and in Appendix~D, 
this theory can be derived from the SU(2) gauge theory above after integrating out the fermionic chargons, and representing $R$ in terms of a bosonic spinon field $z_\alpha$ by
\begin{equation}
R_i = \begin{pmatrix}
z_{i,\uparrow} & -z_{i,\downarrow}^* \\  z_{i,\downarrow} & z_{i,\uparrow}^*
\end{pmatrix}, 
\qquad |z_{i,\uparrow}|^2 + |z_{i,\downarrow}|^2 = 1. \label{SpinonInZ}
\end{equation}
However, integrating out the chargons is only safe when there is a chargon gap, and so the
theories below can compute critical properties of phase transitions only in insulators.

We will not start here from the SU(2) gauge theory, but present a direct derivation
from earlier analyses of the quantum fluctuations of a $S=1/2$ square lattice antiferromagnet near a N\'eel state,
which obtained the following action \cite{SJ90} for a $\mathbb{CP}^1$ theory 
over two-dimensional space ($r=(x,y)$) and time ($t$)
\beq
\mathcal{S} = \frac{1}{g} \int d^2 r d t \, |(\partial_\mu - i a_\mu) z_\alpha|^2 + \mathcal{S}_B. \label{CP1}
\eeq
Here $\mu$ runs over 3 spacetime components, and $a_\mu$ is an emergent U(1) gauge field. 
The local N\'eel order ${\bm n}$ is related to the $z_\alpha$ by ${\bm n} = z_\alpha^\ast {\bm \sigma}_{\alpha\beta} z_\beta^{\vphantom \ast}$ where ${\bm \sigma}$ are the Pauli matrices.
The U(1) gauge flux is defined modulo $2 \pi$, and so the gauge field is
compact and monopole configurations with total flux $2\pi$ are permitted in the path integral. The continuum action in Eq.~(\ref{CP1}) should
be regularized to allow such monopoles. $\mathcal{S}_B$ is the Berry phase of the monopoles \cite{Haldane88,NRSS89,NRSS90}. Monopoles  are suppressed in the states with $\mathbb{Z}_2$ topological order \cite{NRSS91,SSNR91}, and so we do not display the explicit
form of $\mathcal{S}_B$.

The phases of the $\mathbb{CP}^1$ theory in Eq.~(\ref{CP1}) have been extensively studied. For small $g$,
we have the conventional N\'eel state, D$^\prime$ in Fig.~\ref{fig:pdiags}a, with $\langle z_\alpha \rangle \neq 0$ and $\langle {\bm n} \rangle \neq 0$.
For large $g$, the $z_\alpha$ are gapped, and the confinement in the compact U(1) gauge theory leads
to valence bond solid (VBS) order \cite{NRSS89,NRSS90}, which is phase D in Fig.~\ref{fig:pdiags}b. A deconfined critical theory describes the 
transition between these phases \cite{senthil1}.

We now want to extend the theory in Eq.~(\ref{CP1}) to avoid confinement and obtain states with topological order.
In a compact U(1) gauge theory, condensing a Higgs field with charge 2 leads to a phase with deconfined $\mathbb{Z}_2$ charges \cite{FradkinShenker}.
Such a deconfined phase has the $\mathbb{Z}_2$ topological order \cite{NRSS91,SSNR91,Wen91,Bais92,MMS01,Hansson04} of interest to us here.
So we search for candidate Higgs fields with charge 2, composed of pairs of long-wavelength spinons, $z_\alpha$. We also require the Higgs
field to be spin rotation invariant, because we want the $\mathbb{Z}_2$ topological order to persist in phases without magnetic order.
The simplest candidate without spacetime gradients, $\varepsilon_{\alpha\beta} z_\alpha z_\beta$ (where $\varepsilon_{\alpha \beta}$ is the unit
antisymmetric tensor) vanishes identically. Therefore, we are led to the following Higgs candidates with a single gradient ($a=x,y$)
\beq
P \sim \varepsilon_{\alpha\beta} z_\alpha
\partial_t z_\beta  \quad, \quad   Q_a \sim \varepsilon_{\alpha\beta} z_\alpha
\partial_a z_\beta  \,. \label{PQ}
\eeq
These Higgs fields have been considered separately before. Condensing $Q_a$ was the route to $\mathbb{Z}_2$ topological order
in Ref.~\onlinecite{NRSS91}, while $P$ appeared more recently in Ref.~\onlinecite{YangWang}. 

The effective action for these Higgs fields, and the properties of
the Higgs phases, follow straightforwardly from their transformations under the square lattice space group and time-reversal: we collect these in Table~\ref{table:cp1}.
\begin{table}
\begin{tabular}{|c||c|c|c|c|}
\hline
~~ & ~~~$\mathcal{T}$~~~ & ~~~$T_x$~~~ & ~~$I_x$~~ & ~~$R_{\pi/2}$~~ \\
\hline\hline
~~$z_\alpha$~~ & $\varepsilon_{\alpha\beta} z_\beta$ & $\varepsilon_{\alpha\beta} z_\beta^\ast$ & $z_\alpha$ & $z_\alpha$ \\
\hline
~~$Q_x$~~ & $Q_x $ & $Q_x^\ast$ & $-Q_x$ & $Q_y$ \\
\hline
~~$Q_y$~~ & $Q_y $ & $Q_y^\ast$  & $Q_y$ & $-Q_x$ \\
\hline
~~$P$~~ & $-P $ & $P^\ast$ & $P$ & $P$ \\
\hline
\end{tabular}~\\~\\
\caption{Symmetry signatures of various fields under time reversal ($\mathcal{T}$), translation by a lattice spacing along $x$
($T_x$), reflection about a lattice site with $x \rightarrow -x$, $y \rightarrow y$ ($I_x$), and rotation by $\pi/2$ about
a lattice site with $x \rightarrow y$, $y \rightarrow -x$ ($R_{\pi/2}$).}
\label{table:cp1}
\end{table}
From these transformations, we can add to the action
$\mathcal{S} \rightarrow \mathcal{S} + \int d^2 r d t \, \mathcal{L}_{P,Q}$
\bea
\mathcal{L}_{P,Q} &=&   |(\partial_\mu - 2 i a_\mu) P|^2 +  |(\partial_\mu - 2 i a_\mu) Q_a |^2  \label{LH}
 \\ &+& \lambda_1 P^\ast \, \varepsilon_{\alpha\beta} z_\alpha
\partial_t z_\beta +  \lambda_2 Q_a^\ast  \varepsilon_{\alpha\beta} z_\alpha
\partial_a z_\beta  + \mbox{H.c.} \nn &-& s_1 |P|^2 - s_2 |Q_a|^2 - u_1 |P|^4 - u_2 |Q_a|^4 \,,
 + \ldots  \nonumber
\eea
where we do not display other quartic and higher order terms in the Higgs potential.

For large $g$, we have $\langle z_\alpha \rangle = 0$, and can then determine the spin liquid phases by minimizing the Higgs potential as a function of $s_1$ and $s_2$. When there is no Higgs condensate, we noted earlier that we obtain phase D in Fig.~\ref{fig:pdiags}b.
Fig.~\ref{fig:pdiags}b also indicates that the phases A,B,C are obtained when one or both of the $P$ and $Q_a$ condensates are present. This is justified in Appendix~D by a computation of the quadratic effective action for the $z_\alpha$ from the SU(2) gauge theory: we find just the terms with linear temporal
and/or spatial derivatives as would be expected
from the presence of $P$ and/or $Q_a$ condensates in $\mathcal{L}_{P,Q}$. 

We can confirm this identification  from the symmetry transformations in Table~\ref{table:cp1}:\\
(A) There is only a $P$ condensate, and the gauge-invariant quantity $|P|^2$ is invariant under all symmetry operations. Consequently this is a $\mathbb{Z}_2$ spin liquid with no broken symmetries; it has been previously studied by Yang and Wang \cite{YangWang} using bosonic spinons.\\
(B) With a $Q_a$ condensate, one of the two gauge-invariant quantities $|Q_x|^2 - |Q_y|^2$
or $Q_x^\ast Q_y^{\vphantom \ast} + Q_x^{\vphantom \ast} Q_y^\ast $ must have a non-zero expectation value. Table~\ref{table:cp1} shows that these imply Ising-nematic order, as described previously \cite{NRSS91,SSNR91,CQSS16}. We also require $\langle Q_x^{\vphantom \ast} \rangle \langle Q_y^\ast \rangle$ to be real to avoid breaking translational symmetry. \\
(C) With both and $P$ and $Q_a$ condensates non-zero we can define the gauge invariant order parameter $O_a = P Q_a^\ast + P^\ast Q_a$
(again $\langle P \rangle \langle Q_a^\ast \rangle$ should be real to avoid translational symmetry
breaking). The symmetry transformations of $O_a$ show that it is precisely the `current-loop' order parameter of Ref.~\onlinecite{SSSC17}: it is odd under reflection and time-reversal but not their product. 

A similar analysis can be carried out at small $g$, where $z_\alpha$ condenses and breaks spin rotation symmetry. 
The structure of the condensate is determined by the eignmodes of the $z_\alpha$ dispersion in the A,B,C,D phases,
and this determines that the corresponding magnetically ordered states are precisely A$^\prime$,B$^\prime$,C$^\prime$,D$^\prime$, as in Fig.~\ref{fig:pdiags}a.

We have shown here that a class of topological orders intertwine with the observed broken discrete symmetries in the pseudogap phase of the hole doped cuprates. Precisely these topological orders emerge from a theory of quantum fluctuations of magnetically ordered states obtained by four different methods:
the frustrated classical antiferromagnet, the semiclassical non-linear sigma model, the spin density wave theory, and the $\mathbb{CP}^1$ theory supplemented by the Higgs fields obtained by pairing spinons at long wavelengths. The intertwining of topological order and symmetries can explain why the symmetries are restored when the pseudogap in the fermion spectrum disappears at large doping.

We
thank A.~Chubukov, A.~Eberlein, D.~Hsieh, Yin-Chen He, B. Keimer, T.~V.~Raziman, T.~Senthil, and A.~Thomson for useful discussions.
This research was supported by the NSF under Grant DMR-1360789 and the MURI grant W911NF-14-1-0003 from ARO. Research at Perimeter Institute is supported by the Government of Canada through Industry Canada and by the Province of Ontario through the Ministry of Research and Innovation. SS also acknowledges support from Cenovus Energy at Perimeter Institute. 
MS acknowledges support from the German National Academy of Sciences Leopoldina through grant LPDS 2016-12.

\widetext
\appendix


\section{O(3) non-linear sigma model}

We examined a semi-classical O(3) non-linear sigma model of quantum fluctuations of $\mathcal{H}_J$, which expresses $\hat{\bm S}_i$
in terms of the N\'eel field ${\bm n} (r, t)$
and the canonically conjugate uniform magnetization density ${\bm L} (r, t)$ 
\bea
\hat{\bm S}_i &=& S \eta_i {\bm n}_i \sqrt{ 1 - {\bm L}_i^2/S^2} + {\bm L}_i  \label{nls1} \\
{\bm n}^2 &=& 1 \quad , \quad {\bm n} \cdot {\bm L} = 0\,, \label{constraints}
\eea
where $\eta_i = \pm 1$ on the two sublattices, ${\bm n}_i \equiv {\bm n}(r_i, t)$ and similarly for ${\bm L}_i$. Inserting Eq.~(\ref{nls1}) into Eq.~(\ref{spinH}),
and performing an expansion to fourth order in spatial gradients and powers of ${\bm L}$, we obtain $\mathcal{H}_J = \int d^2 r\, \overline{\mathcal{H}}_J$ (the lattice
spacing has been set to unity):
\bea 
\overline{\mathcal{H}}_J &=& \frac{S^2 (J_1 - 2 J_2 - 4 J_3 + 10 J_4 )}{2} \left[(\partial_x {\bm n})^2 + (\partial_y {\bm n})^2 \right] \nn &~&+ 4( J_1 + 2 J_4 - 4 K S^2) {\bm L}^2 \nn
&~& - \frac{(J_1 - 2 J_2 - 4 J_3 + 10 J_4 - 8 KS^2)}{2} {\bm L}^2 \left[(\partial_x {\bm n})^2 + (\partial_y {\bm n})^2 \right] \nn
&~& - \frac{S^2(J_1 - 2J_2 - 16 J_3 + 34 J_4)}{24} \left[ (\partial_x^2 {\bm n})^2 + (\partial_y^2 {\bm n})^2 \right]
\nn
&~& - \frac{(J_1 + 2 J_2 + 4 J_3 + 10 J_4 -8 K S^2)}{2} \left[(\partial_x {\bm L})^2 + (\partial_y {\bm L})^2 \right] \label{HigherOrderTerms} \\
&~& + \frac{S^2 (J_2 - 8 J_4 - 2 KS^2) }{2} (\partial_x^2 {\bm n}) \cdot (\partial_y^2 {\bm n}) \nn
&~& - 8K S^2 \left[ ({\bm L} \cdot \partial_x {\bm n})^2 + ({\bm L} \cdot \partial_y {\bm n})^2 \right] \nn
&~& - K S^4 [(\partial_x {\bm n}).(\partial_x {\bm n})] [(\partial_y {\bm n}).(\partial_y {\bm n})] \nn
&~& + 2 K S^4 [(\partial_x {\bm n}).(\partial_y {\bm n})]^2 + 16K [{\bm L}^2]^2 \,.  \nonumber
\eea
It is useful to extract the terms important for identifying the phases
\begin{equation}
\overline{\mathcal{H}}_J = \frac{\rho_s}{2} (\partial_a {\bm n})^2 + \frac{1}{2 \chi_\perp} {\bm L}^2 + C_1 ({\bm L}^2)^2 + C_2 (\partial_a {\bm L})^2 + \ldots \,; \nonumber
\end{equation}
In this expression, the stiffness of the N\'eel order is $\rho_s$, and
$\chi_\perp$ is the uniform susceptibility transverse to the local N\'eel order. The coefficients are
\bea
&& \rho_s = (J_1 - 2 J_2 - 4 J_3 + 10 J_4)S^2 \nn
&& \chi_\perp^{-1} = 8(J_1 + 2 J_4 - 4 K S^2)
 \\
&& C_1 = 16K  \,, \quad C_2 = - \frac{(J_1 + 2 J_2 + 4 J_3 + 10 J_4 -8 K S^2)}{2} \,.
\nonumber
\eea
The quantum fluctuations of the spin $S$ antiferromagnet are then described by the
action \cite{ssbook}
\beq
\mathcal{S}_{\bm n} = \int dt  d^2 r \, \Bigl[ {\bm L} \cdot \left( {\bm n} \times \partial_t {\bm n} \right) -  \overline{\mathcal{H}}_J \Bigr] + \mathcal{S}_B \label{Sn}
\eeq
where $\mathcal{S}_B$ is as in Eq.~(\ref{CP1}) but now
associated with `hedgehog' defects in ${\bm n}$ \cite{Haldane88,NRSS89,NRSS90}.

The theory $\mathcal{S}_{\bm n}$ with only the first two terms in $\overline{\mathcal{H}}$ is the same \cite{NRSS90} as the original
$\mathbb{CP}^1$ model in Eq.~(\ref{CP1}), and so displays the phases D$^\prime$ (N\'eel) and D (VBS). Now consider the transition
from D$^\prime$ to the spiral phase B$^\prime$: this occurs when increasing $J_{2,3}$ turns $\rho_s$ negative, and we enter a state
with $\langle \partial_a {\bm n} \rangle$ non-zero and spatially precessing; the pitch of the spiral is determined by 
higher order terms in Eq.~(\ref{HigherOrderTerms}).
Similarly, we transition from state D$^\prime$ to the canted state A$^\prime$ when $\chi_\perp^{-1}$ turns negative with increasing $K$: the
state A$^\prime$ has $\langle {\bm L} \rangle \neq 0$, with a value stabilized by the quartic term $C_1$. Finally, the state C$^\prime$ has both $\langle \partial_a {\bm n} \rangle \neq 0$ and $\langle {\bm L} \rangle \neq 0$, and the second constraint in Eq.~(\ref{constraints}) and $C_2>0$ lead to a conical spiral.

These considerations on the O(3) model can be connected to the $\mathbb{CP}^1$
analysis by the important identity (which follows from ${\bm n} = z_\alpha^\ast {\bm \sigma}_{\alpha\beta} z_\beta^{\vphantom \ast}$ and $ |z_\alpha|^2 =1$)
\begin{equation}
(\partial_\mu {\bm n}) \cdot (\partial_\nu {\bm n}) = 2 ( \varepsilon_{\alpha\beta} z_\alpha^{\vphantom \ast}
\partial_\mu z_\beta^{\vphantom \ast}) (\varepsilon_{\gamma\delta} z_\gamma^\ast \partial_\nu z_\delta^\ast) + \mbox{c.c.} \,.
\label{O3CP1}
\end{equation}
From Eq.~(\ref{PQ}) we therefore have the correspondence
\bea
&& (\partial_a {\bm n}) \cdot (\partial_b {\bm n}) \sim Q_a^\ast Q_b^{\vphantom \ast} + Q_a^{\vphantom \ast} Q_b^\ast \quad , \quad 
(\partial_t {\bm n}) \cdot (\partial_t {\bm n}) \sim |P|^2 \nn
&& \quad\quad\quad\quad\quad\quad (\partial_a {\bm n}) \cdot (\partial_t {\bm n}) \sim Q_a^\ast P + Q_a P^\ast \,. \label{nPQ}
\eea
Using also  $\partial_t {\bm n} \sim 
{\bm n} \times {\bm L}$ (from Eq.~(\ref{Sn})), we can now see that the identifications, in the previous paragraph, of the condensates in the O(3)
model correspond to those of the $\mathbb{CP}^1$ model in Fig.~\ref{fig:pdiags}a. The O(3) model analysis has located the phases of Fig.~\ref{fig:pdiags}a in the parameter space of the lattice model $\mathcal{H}$, and we can also use it to estimate couplings in the $\mathbb{CP}^1$ theory.

\section{Spin density wave theory}
In this appendix, we study the fermionic Hubbard model on the square lattice using a mean-field approach, and show that metallic phases with all four spin-density wave orders discussed in the main text show up close to half-filling. We start with the Hubbard Hamiltonian for the electrons $c_{i, \alpha}$.
\begin{equation}
\mathcal{H}_U = - \sum_{i<j, \alpha} t_{ij} c_{i,\alpha}^\dagger c_{j,\alpha}^{\vphantom\dagger}
-\mu \sum_{i, \alpha} c_{i,\alpha}^\dagger c_{i, \alpha}^{\vphantom\dagger}
+  U \sum_i \hat{n}_{i, \uparrow} \hat{n}_{i, \downarrow}
\end{equation}
where $\alpha$ is a spin-index, $t_{ij} = t_p$ are the hopping parameters for $p$'th nearest neighbors with $t_p \neq 0$ for $p = 1,2,3$ and $4$, $U$ is the Hubbard on-site repulsion and $\mu$ is the chemical potential. We perform a mean-field decoupling of the interaction term as follows \cite{Littlewood91,Strinati91,Dzierzawa1992,IrkhinPRB}:
\bea
U \hat{n}_{i, \uparrow} \hat{n}_{i, \downarrow} = \frac{U}{4} \hat{n}_i^2 - U (\hat{\vec{S}}_i \cdot \vec{u}_i)^2  \rightarrow  - \zeta_i \hat{n}_i - \vec{h}_i \cdot\hat{\vec{S}}_i - \frac{U}{4} \langle \hat{n}_i \rangle^2 + U \langle \hat{\vec{S}}_i \rangle^2
\eea
where $ \hat{\vec{S}}_i = \frac{1}{2}  c^\dagger_{i, \alpha} \bm{\sigma}_{\alpha \beta} c_{i, \beta} $ is the electron spin operator, $\hat{n}_i = \sum_{\alpha} \hat{n}_{i, \alpha}$ is the particle number operator at site $\bm{r}_i$, $\vec{u}_i$ is the unit-vector along the spin-quantization axis, $\zeta_i =  - \frac{U}{2} \langle \hat{n}_{i} \rangle$ is a renormalization of the chemical potential which is henceforth absorbed in $\mu$, and $\vec{h}_i = 2 U \langle \hat{\vec{S}}_i \rangle$ is the mean magnetic field at site $\vec{r}_i$.

We consider states which are translation invariant in the charge sector. Therefore the charge density $\langle \hat{n}_i \rangle  = n$ is the same on every site. We include the possibility of in-plane N\'{e}el and spiral order, as well as ferromagnetic canting in the orthogonal ($z$) direction:
\bea
\left\langle \hat{\bm S}_i \right\rangle = N_0 \left[ \cos \left( \vec{K} \cdot \vec{r}_i \right) \cos(\theta) \, \hat{\bm e}_x + 
\sin \left( \vec{K} \cdot \vec{r}_i \right) \cos(\theta) \, \hat{\bm e}_y +  \sin (\theta)\, \hat{\bm e}_z\right].
\label{eq:spinansatz}
\eea
We expect that having the largest possible magnetization at each site will be energetically more favorable. Therefore, we have neglected the possibility of the collinear incommensurate state (stripes) as that leads to a variation in particle number density. While the N\'{e}el or spiral spin-density wave states can consistently explain the drop in Hall number and longitudinal conductivities in the cuprates \cite{Storey_EPL,EMSH16,CSE17}, stripes seem to be inconsistent with the experimental data \cite{charlebois2017hall}. This provide additional motivation for restricting our study to the states described by Eq.~(\ref{eq:spinansatz}). The assumption of uniform charge density also rules out phase separation into hole-rich and particle-rich regions, which are often found in such mean-field treatments \cite{IrkhinPRB}. In principle, farther interactions beyond a single on-site Hubbard repulsion can help avoid phase separation, but such physics cannot be captured by a mean-field treatment. Finally, we also do not consider possible superconductivity since we are interested in metallic phases (which appear at temperatures above the exponentially small superconducting $T_c$). 

The mean-field grand canonical Hamiltonian can then be written in terms of a $2$-component spinor $\Psi_{\vec{k}}$, the dispersion $\xi_{\k} = - \sum_{j \neq i} t_{ij} e^{i \k \cdot (\vec{r}_{i} - \vec{r}_j )} - \mu $, and $h = |\vec{h}_i| = 2 U N_0$ as
\beq
\mathcal{H}^{MF}_U= \sum_\k C_\k^\dagger h_\k C_\k, \text{ where } h_\k = \begin{pmatrix}
\xi_{\k} - \frac{h }{2} \sin \theta & - \frac{h}{2}\cos \theta \\
- \frac{h}{2}\cos \theta & \xi_{\k + \vec{K}} +  \frac{h }{2} \sin \theta
\end{pmatrix},  \text{ and } C_{\k} = \begin{pmatrix}
c_{\k, \uparrow} \\
c_{\k + \vec{K} ,\downarrow}
\end{pmatrix} .
\eeq
This can be diagonalized by a unitary transformation
\beq
\begin{pmatrix}
c_{\k, \uparrow} \\
c_{\k + \Q, \downarrow}
\end{pmatrix} = \begin{pmatrix}
\cos \phi_\k & \sin \phi_\k \\
- \sin \phi_\k & \cos \phi_\k 
\end{pmatrix} \begin{pmatrix}
\alpha_{\k} \\
\beta_{\k}
\end{pmatrix}, \text{ where } \tan(2 \phi_\k) = \frac{h \cos \theta}{\xi_{\k} - \xi_{\k + \Q} - h \sin \theta}
\eeq
The energies of the upper and lower Hubbard bands are given by ($ s = \pm $)
\beq
E_{\k, s} = \frac{1}{2} \left( \xi_{\k} + \xi_{\k + \Q} + s \; \sqrt{(\xi_{\k} - \xi_{\k + \Q} - h \sin \theta)^2 + h^2 \cos^2 \theta} \right).
\eeq
The free energy of the system in the canonical ensemble is given in the continuum limit by (setting $N_s$ to be the number of lattice sites)
\beq
\frac{E^{MF}}{N_s} =  \sum_{s= \pm} \int \frac{d^2k}{(2\pi)^2} \, E_{\k,s} \, n_F(E_{\k,s}) + \mu n - \frac{U n^2}{4} + \frac{h^2}{4U}, \nn \text{ where } n = \langle \hat{n}_i \rangle =  \sum_{s= \pm} \int \frac{d^2k}{(2\pi)^2} n_F(E_{\k,s}).
\eeq
We first tune $\mu$ to adjust the electron-filling $n$. At a fixed filling, we minimize the mean-field free energy $E^{MF}(h, \theta, \vec{K})$. The values of these parameters at the minima in turn describe the magnetically ordered (or paramagnetic) phase for a given set of hopping parameters $t_{p}$ and Hubbard repulsion $U$.

\begin{figure*}
\begin{center}
\includegraphics[width=2.5in]{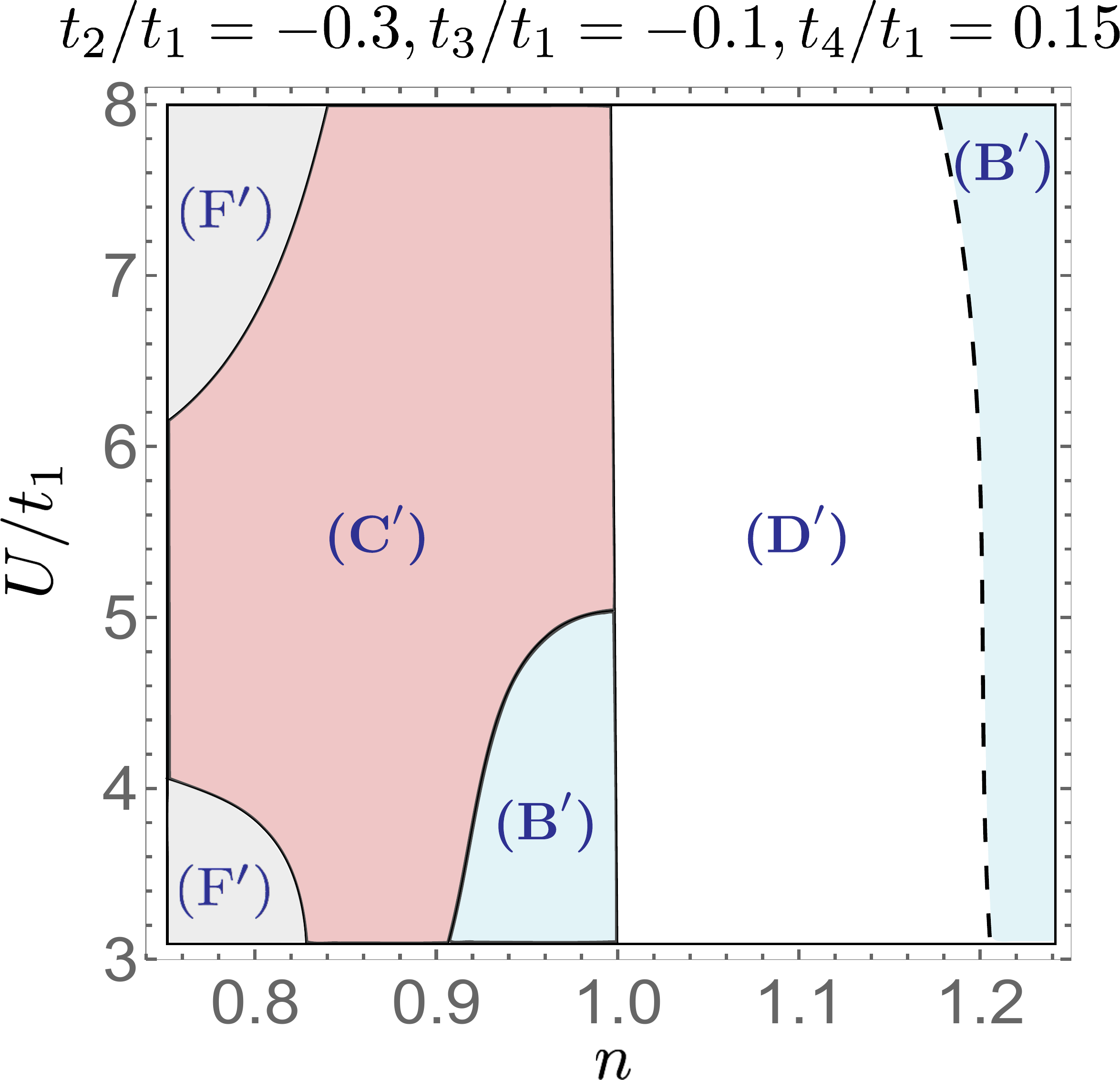}
\end{center}
\caption{Phase diagram from the spin-density wave analysis as a function of Hubbard $U$ and doping $n$ at fixed hopping, showing the N\'{e}el (D$^\prime$), spiral (B$^\prime$), conical spiral (C$^\prime$) and ferromagnetic (F$^\prime$) phases. As in Fig.~\ref{fig:magorder}, solid (dashed) lines are used to represent second (first) order transitions.}
\label{fig:magorder2}
\end{figure*}

As shown in Fig.~\ref{fig:magorder} in the main text and in Fig.~\ref{fig:magorder2} in this appendix, at large $U$ we find exactly the 4 kinds of spin-density wave phases (D$^\prime$) $\vec{K}= (\pi, \pi)$, $\theta=0$, (A$^\prime$) $\vec{K}= (\pi, \pi)$, $0<\theta<\pi/2$,
(B$^\prime$) $\vec{K}$ incommensurate, $\theta=0$, and  (C$^\prime$) $\vec{K}$ incommensurate, $0<\theta<\pi/2$. As expected, in the insulator ($n = 1$) the nearest neighbor Heisenberg exchange $J_1$ is dominant at large $U$, and we find the insulator to be always in the  N\'{e}el phase (D$^\prime$). In the metallic states, the N\'{e}el phase only appears close to zero doping, while the other three antiferromagnetic phases appear contiguous to the N\'{e}el phase. The presence of $t_p$ for $p > 1$ breaks particle-hole symmetry. It is interesting to note that the canted phases appear only on the hole-doped side ($n < 1$) while the electron-doped side has coplanar magnetic order (even at larger dopings not shown in Figs.~\ref{fig:magorder} and \ref{fig:magorder2}). Finally, an additional ferromagnetic phase ($F^\prime$) with $\theta = \pi/2$ also shows up at low enough hole-doping, consistent with previous mean-field studies of the Hubbard model \cite{Dzierzawa1992,IrkhinPRB}.

\section{SU(2) gauge theory}
In this appendix, we derive the effective chargon Hamiltonian (\ref{ChargonHam}) from the Hubbard model in Eq.~(\ref{HubbardModel}) and study the symmetries, together with the associated current and bond patterns, of the different Higgs-field configurations stated in the main text.

\subsection{Effective chargon Hamiltonian}

We write the Hubbard Hamiltonian $\mathcal{H}_U$ as a coherent state path integral and decouple the interaction using a Hubbard-Stratonovich field $\vec{\Phi}_i$. This yields the equivalent action $\mathcal{S}=\mathcal{S}_c+ \mathcal{S}_{\text{int}} + \mathcal{S}_\Phi$, where ($\beta$ and $\tau$ denote inverse temperature and imaginary time, respectively) 
\begin{equation}
\mathcal{S}_c = \int_0^\beta  \diff\tau\left[\sum_{i,\alpha} c^\dagger_{i,\alpha}(\partial_\tau -\mu) c^\pdagger_{i,\alpha} - \sum_{i<j,\alpha} t_{ij} c^\dagger_{i,\alpha} c^\pdagger_{j,\alpha}  \right] \label{Sc}
\end{equation}
describes the hopping of the electrons on the square lattice; The electrons are coupled to the Hubbard-Stratonovich field via  
\begin{equation}
\mathcal{S}_{\text{int}} = \int_0^\beta  \diff\tau \sum_i c^\dagger_{i,\alpha} \vec{\sigma}_{\alpha\beta} c^\pdagger_{,i\beta} \cdot \vec{\Phi}_i \label{SpinFermionCoupling}
\end{equation}
and the action of $\vec{\Phi}_i$ reads as
\begin{equation}
\mathcal{S}_\Phi = \frac{3}{2 U} \int_0^\beta  \diff\tau \sum_i \vec{\Phi}_i^2 .\label{SpinAction0}
\end{equation}

We next transform the electrons to a rotating reference frame as defined in \equref{R}. To rewrite the action in terms of the new degrees of freedom, the chargon and spinon fields $\psi_i$ and $R_i$, let us first focus on $\mathcal{S}_c$. The hopping terms assume the form
\begin{align}
 \sum_{\alpha} t_{ij} c^\dagger_{i,\alpha} c^\pdagger_{j,\alpha} =  \sum_{s,s',\beta}t_{ij} \psi^\dagger_{i, s} \bigl(R^\dagger_i\bigr)_{s\beta}\left(R_j\right)_{\beta s'} \psi^\pdagger_{j,s'} \label{RHSofHopp}
\end{align}
using $\alpha,\beta=\uparrow,\downarrow$ and $s,s'=\pm$ as physical spin and SU(2)-gauge indices, respectively.
To make the quartic term accessible analytically, we perform a mean-field decoupling. Upon introducing $\left(U_{ij}\right)_{s s'} = \braket{\bigl(R^\dagger_iR_j\bigr)_{s s'}}$ and $\left(\chi_{ij}\right)_{ss'} =  \braket{\psi^\dagger_{i,s}\psi^\pdagger_{j,s'}}$, \equref{RHSofHopp} becomes
\begin{equation}
t_{ij} \sum_{s,s'}\left( \psi^\dagger_{i,s} \left(U_{ij}\right)_{ss'} \psi^\pdagger_{j,s'} +  \left(\chi_{ij}\right)_{ss'} \bigl(R^\dagger_iR_j\bigr)_{ss'}\right).
\end{equation}
In the same way, we can rewrite and decouple the time-derivative and chemical potential terms in \equref{Sc}. 

Introducing the `Higgs' field $\vec{H}_i$ according to (cf.~\equref{H})
\begin{equation}
\vec{\sigma}\cdot\vec{H}_i = R_i^\dagger \vec{\sigma}R^\pdagger_i\cdot\vec{\Phi}_i,
\end{equation}
the remaining parts of the action, $\mathcal{S}_{\text{int}}$ and $\mathcal{S}_\Phi$, can be restated as
\begin{align}
\mathcal{S}_{\text{int}} &= \int_0^\beta  \diff\tau \,\vec{H}_i \cdot \sum_{i,s,s'} \psi^\dagger_{i,s} \vec{\sigma}_{ss'} \psi^\pdagger_{i,s'}, \\
\mathcal{S}_\Phi &= \frac{3}{2 U} \int_0^\beta  \diff\tau \sum_i \vec{H}_i^2. \label{BareHiggs}
\end{align}

Taken together, the new action consists of three parts: The effective chargon action,
\begin{align}\begin{split}
\mathcal{S}_\psi = \int_0^\beta  \diff\tau\Biggl[&\sum_{i,s} \psi^\dagger_{i,s}(\partial_\tau - \mu) \psi^\pdagger_{i,s} - \sum_{i<j,s,s'} t_{ij} \psi^\dagger_{i,s} \left(U_{ij}\right)_{ss'} \psi^\pdagger_{j,s'} \\ &+  \sum_{i,s,s'} \vec{H}_i \cdot \psi^\dagger_{i,s} \vec{\sigma}_{ss'} \psi^\pdagger_{i,s'} \Biggr], \label{ChargonPart0}
\end{split}\end{align}
the spinon action ($\text{tr}$ denotes the trace in SU(2) space), 
\begin{equation}
\mathcal{S}_R = \int_0^\beta  \diff\tau\,\,\text{tr}\left[\sum_i \chi_{ii}^T R^\dagger_i \partial_\tau R^\pdagger_i - \sum_{i,j} t_{ij} \chi_{ij}^T R^\dagger_iR_j\right], \label{SpinonAction1}
\end{equation}
which will be discussed in detail in Appendix~D below, and the bare Higgs action in \equref{BareHiggs}. 

Let us for now assume that $U_{ij}$ in \equref{ChargonPart0} is trivial in SU(2) space, $\left(U_{ij}\right)_{ss'}=Z_{ij} \delta_{ss'}$. This should be seen as the first step in or the `ansatz' for an iterative self-consistent calculation of $\chi_{ij}$ and $U_{ij}$ where these two quantitities are inserted in and calculated from the spinon and chargon actions until convergence is reached. At the end of this appendix, we will show that there are no qualitative changes when the self-consistent iterations are carried out.

For $\left(U_{ij}\right)_{ss'}=Z_{ij} \delta_{ss'}$, we recover the effective chargon Hamiltonian (\ref{ChargonHam}) stated in the main text. Furthermore, the bare Higgs field action and the coupling of the Higgs to the chargons is mathematically equivalent to the bare action of the Hubbard-Stratonovich field $\vec{\Phi}_i$ and its coupling to the electrons. For this reason, we can directly transfer the results of the spin-density-wave calculation of Appendix~B to the Higgs phase. The main modification is an order-one rescaling of the hopping parameters from the bare electronic values $t_{ij}$ to those of the chargons $Z_{ij}t_{ij}$.

\subsection{Symmetries and current patterns}
Let us next analyze the symmetries of the effective chargon Hamiltonian (\ref{ChargonHam}) for the different Higgs condensates parameterized in \equref{higgsansatz} of the main text. 

As a consequence of the SU(2) gauge redundancy, a lattice symmetry $g$ with real space action $i \rightarrow g(i)$ is preserved if and only if there are SU(2) matrices $G_i(g)$ such that the effective chargon Hamiltonian is invariant under
\begin{equation}
\psi_{i, s} \rightarrow \sum_{s'}\left(G_i(g)\right)_{s,s'} \psi_{g(i) , s'}. \label{GaugeSymmetry}
\end{equation}
To illustrate the nontrivial consequences of the additional gauge degree of freedom, let us consider translation symmetry $g=T_\mu$, $\mu=x,y$, with $T_\mu(i)=i+\hat{\vec{e}}_\mu$. We first note that all configurations in \equref{higgsansatz} satisfy 
\begin{equation}
  \left\langle {\bm H}_{i+\vec{e}_\mu} \right\rangle = \begin{pmatrix} \, \, R(K_\mu)  & \begin{matrix} 0 \\ 0 \end{matrix} \\ \begin{matrix} 0 & 0 \end{matrix} & 1 \end{pmatrix} \left\langle {\bm H}_{i} \right\rangle, \label{TranslationOfHiggs}
\end{equation}  
where $R(\varphi)$ is a $2\times 2$ matrix describing the rotation of 2D vectors by angle $\varphi$.
As the matrix in \equref{TranslationOfHiggs} belongs to SO(3) and the Higgs field transforms under the adjoint representation of SU(2), we can always find $G_i(T_\mu)$ to render the chargon Hamiltonian invariant; Translation symmetry is thus preserved in all Higgs phases discussed in the main text.

To present an example of broken translation symmetry, let us consider the `staggered conical spiral', labeled by $(\text{E})^{(\eta_x,\eta_y)}_{\vec{K}}$, $\eta_\mu = \pm 1$, in the following, where with $\vec{r}_i = (i_x, i_y)$
\begin{equation}
\left\langle {\bm H}_i \right\rangle = H_0 \left[ \cos \left( \vec{K} \cdot \vec{r}_i \right) \cos(\theta) \, \hat{\bm e}_x + \sin \left( \vec{K} \cdot \vec{r}_i \right) \cos(\theta) \, \hat{\bm e}_y + \eta_{x}^{i_x} \eta_{y}^{i_y} \sin (\theta)\, \hat{\bm e}_z\right], \label{HiggsParam}
\end{equation}
with at least one of $\eta_{x}, \eta_{y}$  equal to $-1$, $0<\theta<\pi/2$, and incommensurate $\vec{K}$. We chose this particular example since we have found the associated magnetically ordered phase as the ground state in the classical analysis of the spin model in \equref{spinH}. It is not visible in Fig.~\ref{fig:magorder}(a)--(c) as it only appears for larger values of $J_2$. For this configuration, the $1$ in the matrix in \equref{TranslationOfHiggs} has to be replaced by $\eta_\mu$. If $\eta_\mu=-1$, the matrix in \equref{TranslationOfHiggs} has determinant $-1$ and, hence, does not belong to SO(3). Consequently, translation symmetry along $\mu$ is broken if $\eta_\mu=-1$. Note that $\Theta T_\mu$, with $\Theta$ denoting time-reversal, is still a symmetry since the Higgs field is odd under $\Theta$.

Similarly, time-reversal and all other lattice symmetries of the effective chargon Hamiltonian can be analyzed. The result is summarized in Table~\ref{TableSymmetries} where the residual symmetries of all the phases with $\mathbb{Z}_2$ topological order discussed in the main text are listed. Note that time-reversal-symmetry breaking necessarily requires a non-collinear Higgs phase since, otherwise, $\vec{H}_i \rightarrow -\vec{H}_i$ can be undone by a global gauge transformation (a global rotation of the Higgs field). 

\begin{table}[bt]
\begin{center}
\caption{Generators of the residual symmetry group of the Higgs phases with $\mathbb{Z}_2$ topological order shown in Fig.~\ref{fig:pdiags}b of the main text. We use $\Theta$ to denote time-reversal, $C_n$ for $n$-fold rotation along the $z$ axis. $I_x$ ($I_y$) and $I_\pm$ are the reflections with action $x\rightarrow -x$ ($y \rightarrow -y$) and at the plane spanned by $x=\pm y$ and the $z$ axis, respectively.}
\label{TableSymmetries}
 \begin{tabular} {cc} \hline \hline
 \hspace{1em} Higgs Phase \hspace{1em} & \hspace{1em} Residual generators  \hspace{1em}  \\ \hline
   (A)  & $T_\mu$, $C_4$, $I_y$, $\Theta$    \\
(B)$_{(k,\pi)}$/(B)$_{(\pi,k)}$  & $T_\mu$, $C_2$, $I_{y}$, $\Theta$   \\ 
(B)$_{(k,k)}$/(B)$_{(k,-k)}$  & $T_\mu$, $C_2$, $I_{+}$, $\Theta$   \\ 
(C)$_{(k,\pi)}$/(C)$_{(\pi,k)}$  & $T_\mu$, $\Theta C_2$, $I_{y/x}$    \\ 
(C)$_{(k,k)}$/(C)$_{(k,-k)}$  & $T_\mu$, $\Theta C_2$, $I_{+/-}$    \\\hline \hline
 \end{tabular}
\end{center}
\end{table}

A complementary and physically insightful approach of detecting and visualizing broken symmetries is based on calculating the (time-reversal symmetric) kinetic energies $K_{ij}$ and the (time-reversal odd) currents $J_{ij}$ on the different bonds $(i,j)$ of the lattice in the ground state of the chargon Hamiltonian. 
These two quantities are defined as and calculated from $K_{ij}=-2\text{Re}\,T_{ij}$ and $J_{ij}=2\text{Im}\,T_{ij}$ where
\begin{equation}
T_{ij} = Z_{ij}t_{ij} \sum_{s} \bigl\langle \psi^\dagger_{i,s}\psi^\pdagger_{j,s}  \bigr\rangle_{\langle\vec{H}_{i}\rangle}.
\end{equation}
Here $\langle \dots\rangle_{\langle\vec{H}_{i}\rangle}$ denotes the expectation values with respect to the ground state of the chargon Hamiltonian $\mathcal{H}_\psi$ in \equref{ChargonHam} for a given Higgs condensate $\vec{H}_i \rightarrow \langle\vec{H}_{i}\rangle$. The kinetic energies $K_{ij}$ (black solid and dashed lines) and, if finite, the currents $J_{ij}$ (black arrows) along the different bonds are illustrated in Fig.~\ref{fig:pdiags}(b) for the three different phases (A)--(C) with $\mathbb{Z}_2$ topological order focussing on a model where only the nearest $t_1$ and next-to-nearest neighbor hopping $t_2$ are non-zero.

\begin{figure}
\begin{center}
\includegraphics[width=4.2in]{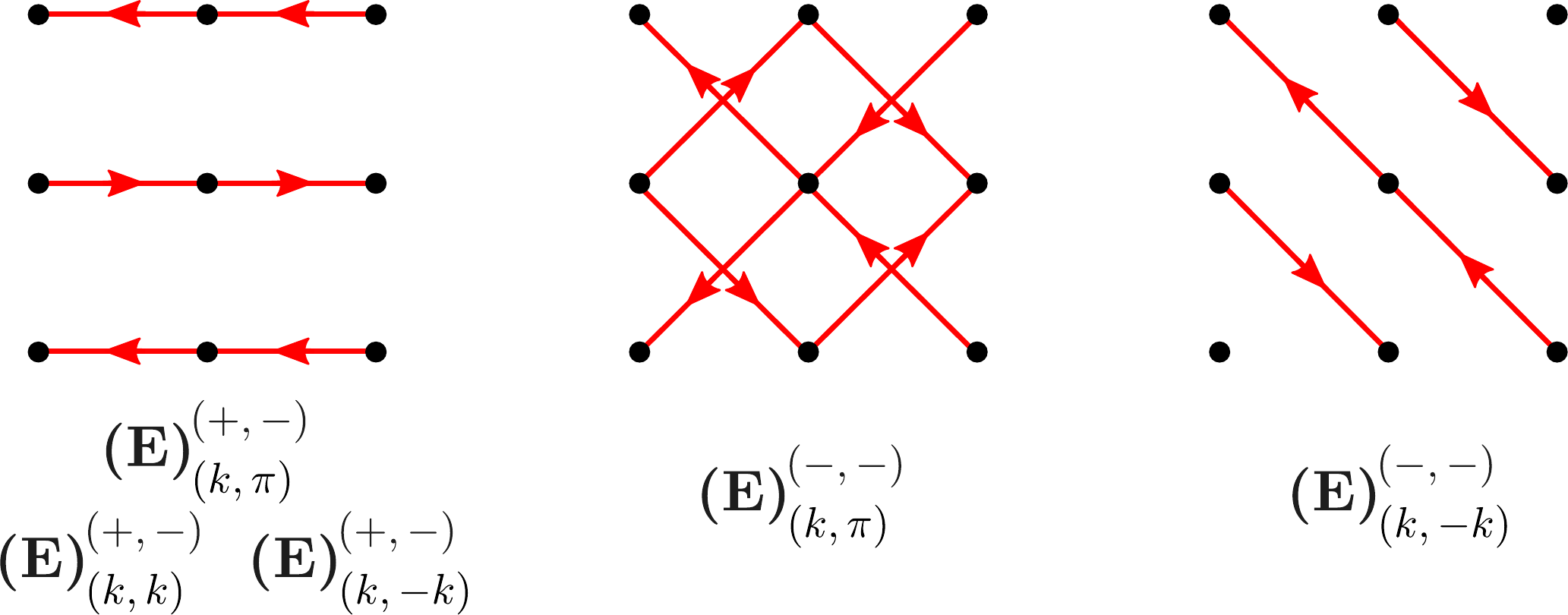}
\end{center}
\caption{Bond currents (arrows) are shown for the different (symmetry inequivalent) staggered conical spiral Higgs field configurations that allow for finite currents in a model with only nearest and next-to-nearest neighbor hopping on the square lattice (dots).}
\label{LoopCurrents}
\end{figure}

For completeness, we also illustrate the current patterns for the staggered conical spiral phases in Fig.~\ref{LoopCurrents}. Here, four unit cells of the square lattice are shown as translation by one lattice site $T_\mu$ is broken if $\eta_\mu=-1$ while $T_\mu^2$ is preserved. 

Three comments on the staggered conical spiral configurations are in order. We first note that the (magnetic) point symmetries of $(\text{E})^{(\eta_x,\eta_y)}_{(k,\pi)}$ and $(\text{E})^{(\eta,\eta)}_{(k,k)}$ are the same as those of $(\text{C})_{(k,\pi)}$ and $(\text{C})_{(k,k)}$, given in Table \ref{TableSymmetries}, while $(\text{E})^{(\eta,-\eta)}_{(k,k)}$ only has $\Theta C_2$ symmetry.
Secondly, the nearest neighbor current operator $J_{i,i+\hat{\vec{e}}_\mu}$ must be zero if $\eta_\mu=-1$ since the residual symmetry $\Theta C_2$ implies $J_{i,i+\hat{\vec{e}}_\mu}=J_{i-\hat{\vec{e}}_\mu,i}$ while $\Theta T_\mu$ leads to $J_{i,i+\hat{\vec{e}}_\mu}=-J_{i-\hat{\vec{e}}_\mu,i}$. For the same reason, we conclude that the diagonal currents must vanish if $\eta_x\eta_y=-1$.
Third, notice that the configuration $(\text{E})^{(+,-)}_{(\pi,k)}$ cannot support finite currents in the model with nearest and next-to-nearest-neighbor hopping since the (magnetic) point symmetries and the (magnetic) translations are only consistent with $J_{ij}=0$ along all bonds of the lattice (Finite currents are possible in the presence of third-nearest-neighbor hopping. The associated current pattern is not shown in Fig.~\ref{LoopCurrents}).

Finally, let us come back to the issue of calculating $U_{ij}$ (and $\chi_{ij}$ entering the spinon action) self-consistently. While we had used the ansatz of diagonal $U_{ij}$ for the iteration, recalculating $U_{ij}$ from the spinon action will in general also yield non-vanishing off-diagonal components; However, the symmetry analysis we discussed above is not affected since symmetries are preserved in the iteration process. To see this, assume that the chargon Hamiltonian $\mathcal{H}_\psi[\vec{H}_i,U_{ij}^{(0)}]$ with a certain Higgs field configuration $\vec{H}_i$ and gauge connection $U_{ij}=U_{ij}^{(0)}$ (\emph{e.g.}, $U_{ij}^{(0)}=Z_{ij}\sigma_0$ in the first iteration) is invariant under a symmetry operation $g$, \emph{i.e.}, invariant under (\ref{GaugeSymmetry}). This implies that the $\chi_{ij}$ calculated from $\mathcal{H}_\psi[\vec{H}_i,U_{ij}^{(0)}]$ satisfy $\chi_{ij}= G^*_i(g)\chi_{g(i)g(j)} G^T_{j}(g)$. Consequently, the spinon action in Eq.~(\ref{SpinonAction1}) is symmetric under $R_i \rightarrow R_{g^{-1}(i)} G_{g^{-1}(i)}(g)$, where $g^{-1}$ denotes the inverse of $g$. The `new' or `updated' gauge connection $U_{ij}=U^{(1)}_{ij}$ as obtained from the spinon action thus satisfies
\begin{equation}
U^{(1)}_{g(i)g(j)} = G^\dagger_{i}(g)U^{(1)}_{ij}    G^\pdagger_{j}(g) 
\end{equation}
and, hence, the `new' chargon Hamiltonian $\mathcal{H}_\psi[\vec{H}_i,U_{ij}^{(1)}]$ is still invariant under (\ref{GaugeSymmetry}). This means that the symmetries and the qualitative form of the bond as well as current patterns discussed above are unaffected by replacing $U_{ij}=U_{ij}^{(0)}=Z_{ij}\sigma_0$ by the, generally non-diagonal, self-consistent solution $U_{ij}$ obtained via iteration. The off-diagonal components in $U_{ij}$ should be seen as additional corrections to the energetics of the spin-density-wave analysis of Appendix~B and, hence, are expected to only lead to small changes in the phase boundaries in Fig.~\ref{fig:magorder}d-f.

\section{Derivation of $\mathbb{CP}^1$ theory from SU(2) gauge theory}
To derive the $\mathbb{CP}^1$ actions for the different phases in Fig.~\ref{fig:pdiags}b from the SU(2) gauge theory, it is convenient to use the gauge where the Higgs field is given by
\begin{equation}
\left\langle {\bm H}_i \right\rangle = (-1)^{i_x + i_y} H_0 \, \hat{\vec{e}}_z, \label{AFMgauge}
 \end{equation} 
\emph{i.e.}, the Higgs field has the form of the spin configuration of an antiferromagnet.
Choosing the `antiferromagnetic gauge' in \equref{AFMgauge} is possible for all configurations in \equref{higgsansatz} or, more generally, in \equref{HiggsParam}, since the Higgs field transforms under the adjoint representation of SU(2) and $|\left\langle {\bm H}_i \right\rangle| = H_0$.

In this gauge, the relation between the $\mathbb{CP}^1$ fields $z_i=(z_{i\uparrow},z_{i\downarrow})$ and the spinons $R_i$ is given by \equref{SpinonInZ} for all Higgs configurations. This is verified by noting that \equref{H} with $\hat{\vec{S}}_i = (-1)^{i_x + i_y} \vec{n_i}$ and $\vec{n_i}= z_i^\dagger \vec{\sigma} z_i^\pdagger$ will hold for $R_i$ given in \equref{SpinonInZ} if the Higgs field has the form (\ref{AFMgauge}).
As $R_i$ transforms nontrivially under SU(2) gauge transformations, see \equref{GaugeTrafo}, the relation between $z_i$ and $R_i$ is generally different in a different gauge.

Inserting the parameterization (\ref{SpinonInZ}) into \equref{SpinonAction1} and writing $\chi_{ij}^{ss'}=(\chi_{ij})_{ss'}$ yields the general form 
\begin{align}\begin{split}
\mathcal{S}_{z} = & \int_0^\beta  \diff\tau \Biggl[\sum_i \Bigl( (\chi_{ii}^{++} -\chi_{ii}^{--}) z_i^\dagger \partial_\tau z^\pdagger_i + \chi_{ii}^{-+} \varepsilon_{\alpha\beta} z_{i,\alpha} \partial_\tau z_{i,\beta}  - \chi_{ii}^{+-} \varepsilon_{\alpha\beta} z^*_{i,\alpha} \partial_\tau z^*_{i,\beta} \Bigr) \\ & -\sum_{i<j} t_{ij} \Bigl( (\chi_{ij}^{++}+\chi_{ji}^{--})z^\dagger_i z_j  + (\chi_{ij}^{-+}-\chi_{ji}^{-+})\varepsilon_{\alpha\beta} z_{i, \alpha}z_{j, \beta} + \text{c.c.}\Bigr)   \Biggr],\label{GenCP}\end{split}\end{align}
of the $\mathbb{CP}^1$ action. We already notice (the lattice form of) the charge-$2$ terms $\varepsilon_{\alpha\beta} z_\alpha \partial_{\tau} z_\beta$ and $\varepsilon_{\alpha\beta} z_\alpha
\partial_a z_\beta$, $a=x,y$, coupling to the Higgs fields $P$ and $Q_a$ in \equref{LH}.

Depending on the symmetries of the chargon Hamiltian, some of the terms in \equref{GenCP} have to vanish as we will discuss next. This corresponds to the absence of condensation of one or both of the Higgs fields $P$ and $Q_a$ in the phases (A), (B), and (D).  

Since this has already been discussed for the case of phase (D) in Ref.~\onlinecite{OurPreprint}, we focus here on the other three cases. To begin with phase (B), we apply the gauge transformation $V_{i} = e^{-i \pi \sigma_y/4} e^{i \delta \vec{K}\cdot \vec{r}_i \sigma_x/2}$, where $\delta\vec{K}=\vec{K}-(\pi,\pi)$, to bring the associated Higgs field configuraton in the form of \equref{AFMgauge}. In the resulting `antiferromagnetic gauge', the chargon Hamiltonian reads as
\begin{equation}
\mathcal{H}^{(B)}_\psi = - \sum_{i<j, s,s'} t_{ij} Z_{ij} \psi_{i,s}^\dagger \left( e^{i \delta\vec{K}\cdot (\vec{r}_j-\vec{r}_i) \frac{\sigma_x}{2}} \right)_{ss'} \psi_{j,s'}^{\vphantom\dagger} 
-\mu \sum_{i, s} \psi_{i,s}^\dagger \psi_{i ,s}^{\vphantom\dagger}
- H_0 \sum_{i,s}  (-1)^{i_x+i_y} s\, \psi_{i,s}^\dagger \psi_{i,s}^{\vphantom\dagger} \,. \label{ChargonHam1}
\end{equation}
We see that $\vec{K}\neq (\pi,\pi)$, \emph{i.e.}, $\delta\vec{K}\neq 0$, is required to have off-diagonal matrix elements in SU(2) space ($s\neq s'$) in the Hamiltonian. These are necessary for charge-$2$ terms in the $\mathbb{CP}^1$ action as otherwise $\chi_{ij}^{s,-s}=0$ and, hence, the prefactors of $\varepsilon_{\alpha\beta} z_{i,\alpha} \partial_\tau z_{i,\beta}$ and $\varepsilon_{\alpha\beta} z_{i, \alpha}z_{j, \beta} $ vanish in \equref{GenCP}.
Alternatively, this can also be seen by noting that $R_i V^\dagger_i = R_i|_{z_i \rightarrow z_i e^{i \xi}}$ for the global gauge transformation $V_i=e^{-i \xi \sigma_z}$ and that all terms in the chargon Hamiltonian $\mathcal{H}^{(B)}_\psi$ are invariant under this gauge transformation, $\psi_i \rightarrow V_i \psi_i$, except for the contributions of finite $\delta\vec{K}$ to the  hopping term. This means that the effective $\mathbb{CP}^1$ action must be invariant under $z_i \rightarrow z_i e^{i \xi}$ in the limit $\delta\vec{K}=0$, \emph{i.e.}, the charge-$2$ terms can only arise if $\delta\vec{K}\neq 0$.

To see which of the two possible charge-2 terms in Eq.~(\ref{GenCP}) can be non-zero, let us analyze the symmetries of the chargon Hamiltonian $\mathcal{H}^{(B)}_\psi $ for non-zero $\delta \vec{K}$. We first note that $\mathcal{H}^{(B)}_\psi $ is invariant under $\psi_{i,s}\rightarrow \psi_{i+\hat{\vec{e}}_\mu,-s}$ leading to
\begin{equation}
\chi_{ij}^{ss'}=\chi_{i+\hat{\vec{e}}_\mu j+\hat{\vec{e}}_\mu}^{-s-s'}=\chi_{i+2\hat{\vec{e}}_\mu j+2\hat{\vec{e}}_\mu}^{ss'} . \label{PseudoTransl}
\end{equation}
From this follows
\begin{equation}
\chi_{ii}^{++} -\chi_{ii}^{--} =\chi_{ii}^{++} -\chi_{i+\hat{\vec{e}}_\mu i+\hat{\vec{e}}_\mu}^{++} =  (-1)^{i_x+i_y} \chi_\tau.
\end{equation}
The constant $\chi_\tau$ can be shown to be real: The Hamiltonian $\mathcal{H}^{(B)}_\psi$ commutes with the antiunitary operator $\widetilde{\Theta}$ defined by $\widetilde{\Theta} \psi_{j,s} \widetilde{\Theta}^\dagger = i \sigma_z \psi_{j,s}$. This implies
\begin{equation}
\chi_{ij}^{ss'} = ss' \left(\chi_{ij}^{ss'}\right)^*= ss' \chi_{ji}^{s's}.
\end{equation}
This not only leads to $\chi_\tau \in \mathbb{R}$, but can also be used to rewrite
\begin{subequations}\begin{align}
\chi_{ij}^{++}+\chi_{ji}^{--} &= \chi_{ij}^{++}+\chi_{i+\hat{\vec{e}}_\mu j+\hat{\vec{e}}_\mu}^{++}= \chi^t_{i-j} \in \mathbb{R}, \\
\chi_{ij}^{-+}-\chi_{ji}^{-+} &= \chi_{ij}^{-+}+\chi_{i+\hat{\vec{e}}_\mu j+\hat{\vec{e}}_\mu}^{-+}=\chi^Q_{i-j}=-\chi^Q_{j-i} \in i\,\mathbb{R},
\end{align}\end{subequations}
where we have also taken advantage of \equref{PseudoTransl}.

We finally consider the symmetry of $\mathcal{H}^{(B)}_\psi$ under the unitary transformation $\psi_{j,s}\rightarrow i\sigma_z\psi_{-j,s}$ which, together with Eq.~(\ref{PseudoTransl}), leads to
\begin{equation}
\chi_{ii}^{+-} = -\chi_{-i-i}^{+-}=-\chi_{ii}^{+-}=0.
\end{equation}
Consequently, the terms $\varepsilon_{\alpha\beta} z_{i,\alpha} \partial_\tau z_{i,\beta}$ and $\varepsilon_{\alpha\beta} z^*_{i,\alpha} \partial_\tau z^*_{i,\beta}$ are absent in Eq.~(\ref{GenCP}) for phase (B). 

Taken together, the $\mathbb{CP}^1$ action $\mathcal{S}_z$ in Eq.~(\ref{GenCP}) assumes the form
\begin{align}\begin{split}
\mathcal{S}^{(B)}_{z} =  \int_0^\beta  \diff\tau \Biggl[\sum_i (-1)^{i_x+i_y} \chi_\tau \, z_i^\dagger \partial_\tau z^\pdagger_i   -\sum_{i<j} t_{ij}  \left(\chi^t_{i-j}\, z^\dagger_i z_j  + \chi^Q_{i-j}\,\varepsilon_{\alpha\beta} z_{i ,\alpha}z_{j, \beta} + \text{c.c.} \right)   \Biggr].\end{split} \label{SzBIntermed} \end{align}
For concreteness, let us focus on nearest-neighbor hopping ($t=t_{i,i+\hat{\vec{e}}_\mu}$) and $\delta Q_x = \delta Q_y$ (corresponding to phase (B)$_{(k,k)}$). 
Using that $\chi^{t,Q}_{\hat{\vec{e}}_x}=\chi^{t,Q}_{\hat{\vec{e}}_y}\equiv i\chi^{Q,t}$, treating the constraint $z_i^\dagger z_i^\pdagger = 1$ on average by introducing the Lagrange multiplier $\lambda$, and rewriting
\begin{equation}
z_{i\alpha} \sim z_\alpha(\vec{r}_i) + (-1)^{i_x+i_y} \pi_{\alpha}(\vec{r}_i),
\end{equation}
where $z(\vec{r})$ and $\pi(\vec{r})$ are assumed to be slowly varying continuum fields, a gradient expansion of Eq.~(\ref{SzBIntermed}) yields ($a$ denotes lattice spacing)
 \begin{align}\begin{split}
\mathcal{S}^{(B)}_{z} \sim \int_0^\beta  \diff\tau \int \frac{\diff^2 r}{a^2}  \Biggl[& \chi_\tau (z^\dagger \partial_\tau \pi + \pi^\dagger \partial_\tau z) + (\lambda - t\chi^t)z^\dagger z + (\lambda + t\chi^t)\pi^\dagger \pi \\ & +t \chi^t a^2 \sum_{\mu=x,y} (\partial_\mu z^\dagger )\partial_\mu z + 2t \chi^Q a\sum_{\mu=x,y}\left( \varepsilon_{\alpha\beta} z_{\alpha}\partial_\mu z_{ \beta} + \text{c.c} \right) \Biggr].
\label{PhaseBCPIntermed}\end{split}\end{align}
In Eq.~(\ref{PhaseBCPIntermed}) spatial derivatives up to second (zeroth) order of $z_\alpha$ ($\pi_\alpha$) are kept as these gives rise to the terms of the $\mathbb{CP}^1$ action we are interested in. Indeed, integrating out the $\pi$ field, we recover the $\mathbb{CP}^1$ theory of the main text with Higgs condensates $\braket{Q_x}=\braket{Q_y}\neq 0$ and $\braket{P}=0$.

In a similar way, the remaining phases, (A) and (C), can be analyzed and one finds the $\mathbb{CP}^1$ action with Higgs condensates summarized in Fig.~\ref{fig:pdiags}b.

\bibliography{pseudogap}

\end{document}